\documentclass[onecolumn,showpacs]{revtex4}

\usepackage{graphicx}% Include figure files
\usepackage{dcolumn}% Align table columns on decimal point
\usepackage{bm}% bold math
\usepackage{color}

\input epsf

\begin{document}

\title{\Large Role of Chameleon Field in Accelerating Universe}

\author{\bf Piyali Bagchi Khatua$^1$\footnote{piyali.bagchi@yahoo.co.in}
and Ujjal Debnath$^2$\footnote{ujjaldebnath@yahoo.com} }

\affiliation{$^1${Department of Computer Science and Engineering,
Netaji Subhas Engineering College, Garia, Kolkata-700 152, India.\\
$^2$Department of Mathematics, Bengal Engineering and Science
University, Shibpur, Howrah-711 103, India.} }

\date{\today}

\begin{abstract}
In this work, we have considered a model of the flat FRW universe
filled with cold dark matter and Chameleon field where the scale
function is taken as, (i) Intermediate Expansion and (ii)
Logamediate Expansion. In the both cases we find the expressions
of Chameleon field, Chameleon potential, statefinder parameters
and slow-roll parameters. Also it has been shown that the
potential is always decreases with the chameleon field in both the
scenarios. The nature of slow-roll parameters have been shown
diagrammatically.
\end{abstract}

\pacs{}

\maketitle

\section{\bf\large{Introduction}}

Dark energy is the mysterious entity that accounts for $70\%$ of
the mass-energy content of the universe and is causing the
expansion of the universe to accelerate. The current cosmological
observations [1-3] and theories suggest that the Universe is
undergoing an accelerated expansion due to dark energy. Possible
candidates for this dark energy are a cosmological constant, a
slowly rolling scalar field, Chaplygin gas, Tachyonic field etc.
The cosmological constant scenario with a spatially flat or a
nearly spatially flat Universe implies an energy density of order
$\rho_{critical} = 3H_{0}^2 /8\pi G$ today. In this model, this
energy density should be constant in time, giving rise to severe
problems (see ref.[4]), making the explanation of dark energy with
a cosmological constant which is unnatural. A more general model
is that of a slowly rolling scalar field, known as quintessence,
which has a negative pressure and therefore accelerates expansion.\\

Researchers at Fermilab in the US have carried out the first
laboratory experiment to look for a hypothetical form of matter
known as Chameleon Particles. Chameleon particles were first
proposed in 2003 by Justin Khoury and Amanda Weltman [5] at
Columbia University as a possible explanation for dark energy. In
places where the density of matter is relatively high, chameleon
particles interact very weakly with other matter and only over
very short distances, which could explain why we have yet to spot
them here on Earth. In the inter-galactic space where matter
density is extremely low, the particles interact much more
strongly with other matter and over very large distances. The
particles could be spotted by how they affect light travelling to
Earth from distant galaxies which are considered details in ref.
[6-8].\\

If chameleon particles could be created within a steel-walled
vacuum chamber of few centimeters in diameter and several meters
long(which is called GammaV Chamber), many of them would be unable
to escape, that is because the mass of a chameleon is proportional
to the local density. Chameleons be created by firing a laser into
a region of the chamber containing a very high magnetic field. If
the chameleon field couples strongly to matter, we would have
already detected it as a fifth force in gravitational experiments.
A candidate one could come up with for such a scalar field could
be the dilation from string theory. There exists a mechanism which
suppresses the coupling of the field to matter, avoiding
detection. Recently, a new model was suggested [5] such that: in
which a scalar field with a thin-shell mechanism couples to matter
with gravitational strength while remaining very light on
cosmological scales. The crucial feature of this field is that the
coupling gives it a mass depending on the local density of matter.
Thus the scalar field is named chameleon. Several works has been
done due to Chameleon field [9-13]. In regions of high matter
density, chameleon field has a large mass and therefore its
interaction with matter is small. In regions of low energy
density, like the solar system, the mass is small and the
interaction should be large and observable. The action of this
field is suppressed by a so called thin-shell mechanism, to be
discussed later. In opposition to slow rolling
scalar field models, chameleons are not limited to quintessence.\\

The paper is organized as follows: In section II, we have
considered a flat FRW model which represents the corresponding
equation of Chameleon Field where the scale function is taken as,
(i) Intermediate Expansion and (ii) Logamediate Expansion. In both
the cases we find the expressions of the Chameleon field,
Chameleon potential and slow-roll parameters. We have taken some
particular values of the parameters and constants for the
graphical representation. The paper ends with a short discussion
in section III.\\

\section{\bf\large{Basic Equations for the Chameleon Field}}

The metric of a spatially flat isotropic and homogeneous Universe
in FRW model is, \begin{equation}
ds^{2}=dt^{2}-a^{2}(t)\left[dr^{2}+r^{2}(d\theta^{2}+sin^{2}\theta
d\phi^{2})\right]
\end{equation}
where $a(t)$ is the scale factor of the universe. The Hubble
parameter is defined as,
\begin{equation}
H = \frac{\dot{a}}{a}
\end{equation}

Consider the relevant action is given by,
\begin{equation} S = \int\sqrt{-g} d^{4}x [ f(\phi){\cal
L}+\frac{1}{2}\phi_{ ,\mu}\phi^{ ,\mu}+\frac{\cal{R}}{16\pi
G}-V(\phi)]
\end{equation}
where, $\phi$ is the {\bf Chameleon Scalar Field} and $V(\phi)$ is
the {\bf Chameleon potential}. Also, ${\cal{R}}$ is the {\bf Ricci
Scalar}, $G$ is the {\bf Newtonian constant} of gravity.
$f(\phi){\cal L}$ is the modified Lagrangian matter and $f(\phi)$
is an analytic function of $\phi$. The cold dark matter and
chameleon field interact with each other. The variation of action
of the above equation w.r.t $\phi$ gives,
\begin{equation}
3H\dot{\phi}+\ddot{\phi}+\frac{dV}{d\phi}+(p+\rho)\frac{df}{d\phi}=0
\end{equation}
which reduces to, \begin{equation}
3H\dot{\phi}^2+\dot{\phi}\ddot{\phi}+\dot{V}+(p+\rho)\dot{f}=0
\end{equation}

which is the wave equation for the chameleon field. Again, the
variation of the same equation w.r.t. metric tensor components
gives (choosing $8\pi G = 1$),
\begin{equation}
3\frac{\dot{a}^2}{a^2}=\rho f +\frac{1}{2} \dot{\phi}^2+V(\phi)
\end{equation}
and \begin{equation} 2\frac{\ddot{a}}{a}+\frac{\dot{a}^2}{a^2}=-pf
-\frac{1}{2}\dot{\phi}^2 +V(\phi)
\end{equation}
which are the corresponding field equations. For simplifying the
above model we consider,
\begin{equation}
V=V_{0}\dot{\phi}^2
\end{equation}
where, $V_{0}$ is a constant, $p$ is the pressure and $\rho$ is
the energy density. Now, combining equations (5), (6) and (7), we
get,
\begin{equation} \frac{\partial}{\partial t}(\rho
f)+3H(p+\rho)f=(p+\rho)\dot{f}
\end{equation}
Simplifying the above equation we get,
\begin{equation}
\frac{d\rho}{p}+3\frac{da}{a}\left(1+\frac{\rho}{p}\right)=\frac{df}{f}
\end{equation}
To get a complete solution of the above differential equation, we
consider the barotropic equation of state
\begin{equation}
p=\omega\rho
\end{equation}
where $\omega$ is a constant. Equations (9) and (11) together
gives,
\begin{equation}
\rho=\rho_{0}a^{-3(1+\omega)}f^\omega
\end{equation}
where $\rho_{0}$ is the integrating constant. If $\omega=0$, then
from (11) and (12) we get,
\begin{equation}
p=0~~~and~~~\rho=\rho_{0}/a^3
\end{equation}
this is the case where a particular matter dominated the universe
and then the fluid is taken in the form of pressureless dust which
is considered in ref [12].

\subsection{\bf\large{Chameleon Field in case of Intermediate
Inflation}}

Consider a particular scenario of {\bf Intermediate Inflation}
[14], where the scale factor $a(t)$ of the Friedmann universe is
described as,
\begin{equation} a(t)=\exp (B t^\beta)
\end{equation}
 where $B\beta>0$, $B>0$ and $0<\beta<1$ are constants.
 The Hubble parameter becomes,
\begin{equation}
H=\frac{\dot{a}}{a}=B\beta t^{\beta-1}
\end{equation}
In this case the expansion of Universe is faster than {\bf
Power-Law Inflation}, where the scale factor is given as,
$a(t)=t^n$, where $n>1$ is a constant. Also, the expansion of the
Universe is slower for {\bf Standard De Sitter Inflation} where
$\beta=1$. Hence we get,
\begin{equation}
\frac{\dot H}{H}=\frac{\beta-1}{t}
\end{equation}
\begin{equation}
\frac{\ddot H}{H}=\frac{(\beta-1)(\beta-2)}{t^2}
\end{equation}
and
\begin{equation}
\frac{\ddot a}{a}=B^2 \beta^2
t^{2\beta-2}+B\beta(\beta-1)t^{\beta-2}
\end{equation}

Putting the value of $a(t)$ the declaration parameter $q=-\frac{a
{\ddot{a}}}{{\dot{a}}^2}$ we get,
\begin{equation}
q=-1-\frac{\beta-1}{B\beta t^\beta}
\end{equation}

Fig.1 represents the variation of $q$ against $H$ for different
values of $\beta$.\\

\begin{figure}
\includegraphics[height=2in]{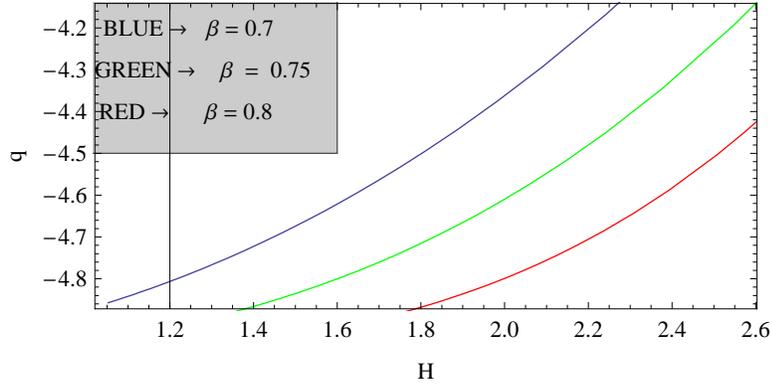}~~~~ \caption{The
variation of $q$ against $H$ from eq. (19) for $B =
3$,$\omega=.2$,$V_{0}$=2 and $\beta=0.7, 0.75, 0.8$} \vspace{7mm}
\end{figure}

The flat Friedmann model which is analyzed in terms of the
statefinder parameters. The trajectories in the $\{s, r\}$ plane
of different cosmological models shows different behavior. The
statefinder diagnostic of SNAP observations used to discriminate
between different dark energy models. The statefinder diagnostic
pair is constructed from the scale factor $a(t)$. The statefinder
diagnostic pair is denoted as $\{s,r\}$ and defined as [15],
\begin{equation}
 r=\frac{\dddot{a}}{a H^3}
 ~~~\text{and}~~~
s=\frac{r-1}{3(q-\frac{1}{2})}
\end{equation}
Using (15) and (19), equation (20) becomes,
\begin{equation}
r=1+\frac{(\beta-1)(\beta-2) }{B^2 \beta^2}t^{-2\beta} +
\frac{\beta+1}{B\beta}t^{-\beta}
\end{equation}
and
\begin{equation}
s=-\frac{\frac{(\beta-1)(\beta-2)}{B\beta
t^\beta}+\beta+1}{3(\beta-1)+\frac{9B\beta t^\beta}{2}}
\end{equation}

Fig.2 represents the variation of $s$ against $r$ for different
values of $\beta$.\\

\begin{figure}
\includegraphics[height=2in]{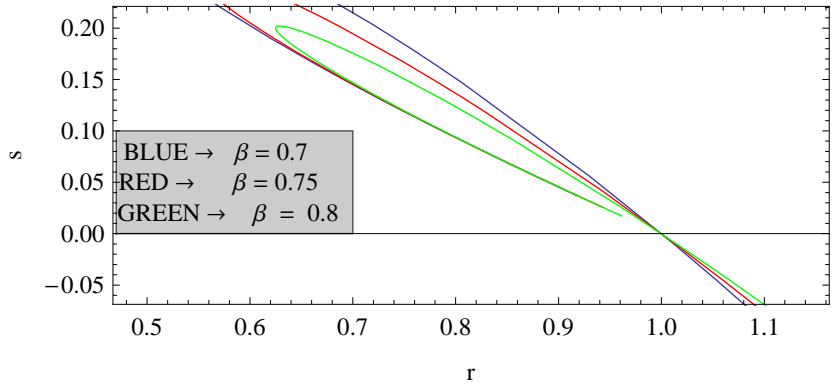}~~~~ \caption{The
variation of $s$ against $r$ from (21) and (22) for $B =
3$,$\omega=.2$,$V_{0}$=2 and $\beta=0.7, 0.75, 0.8$} \vspace{7mm}
\end{figure}

Using (8), (11) and (14), equations (6) and (7) become
\begin{equation}
3B^2 \beta^2 t^{2\beta-2}=\rho f +\frac{2V_{0}+1}{2} \dot{\phi}^2
\end{equation}
and
\begin{equation}
2B^2 \beta^2 t^{2\beta-2}+2B\beta(\beta-1)t^{\beta-2}+B^2 \beta^2
t^{2\beta-2}=-\rho\omega f +\frac{2V_{0}-1}{2}\dot{\phi}^2
\end{equation}
Eliminating $f$ from (23) and (24) we get,
\begin{equation}
3B^2 \beta^2(\omega+1)
t^{2\beta-2}+2B\beta(\beta-1)t^{\beta-2}=\frac{2(\omega+1)V_{0}+(\omega-1)}{2}\dot{\phi}^2
\end{equation}
Thus using (8) and (25) we get the Chameleon Field and Chameleon
Potential as,
\begin{equation}
\phi=\int\sqrt{\frac{3B^2 \beta^2(\omega+1)
t^{2\beta-2}+2B\beta(\beta-1)t^{\beta-2}}{(\omega+1)V_{0}+\frac{\omega-1}{2}}}dt
\end{equation}
and
\begin{equation} V(\phi)=\frac{3B^2 \beta^2(\omega+1)
t^{2\beta-2}+2B\beta(\beta-1)t^{\beta-2}}{\omega+1+\frac{\omega-1}{2V_{0}}}
\end{equation}

Fig.3 ~represents the variation of $V$ against $\phi$ for
different values of $\beta$. It has been seen that the potential
is always decreases with the chameleon field $\phi$.\\

\begin{figure}
\includegraphics[height=3in]{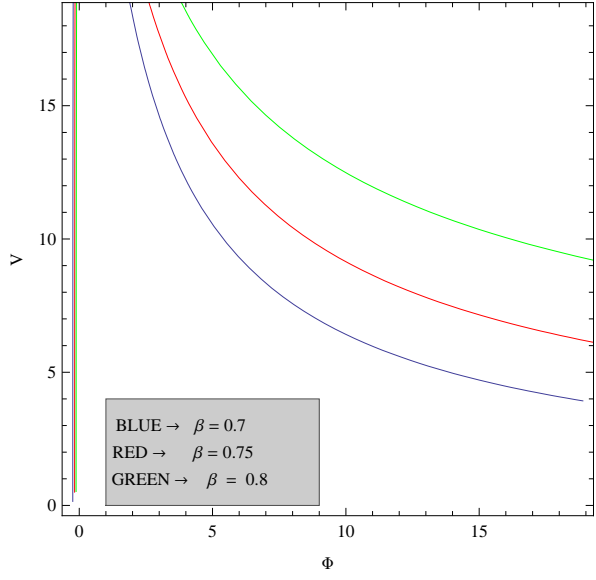}~~~~ \caption{The variation
of $V$ against $\phi$ from (26) and (27) for $B =
3$,$\omega=0.2$,$V_{0}$=2 and $\beta=0.7,0.75,0.8$} \vspace{7mm}
\end{figure}

The slow-roll parameters are defined as [14],
\begin{equation}
\epsilon=2\left(\frac{H'}{H}\right)^2=\frac{2{\dot
H}^2}{H^2\phi^{2}}
\end{equation}
and
\begin{equation}
\eta=\frac{2H''}{H}=\frac{2}{H}\left(\frac{\ddot
H}{\dot\phi^2}-\frac{\dot H \ddot\phi}{\dot\phi^3}\right)
\end{equation}

From (26) we get,
\begin{equation}
\dot{\phi}^2=\frac{3B^2 \beta^2(\omega+1)
t^{2\beta-2}+2B\beta(\beta-1)t^{\beta-2}}{(\omega+1)V_{0}+\frac{\omega-1}{2}}
\end{equation}
Differentiate the above equation w.r.t. $t$ we get,
\begin{equation}
\dot{\phi}\ddot{\phi}=\frac{3B^2 \beta^2(\beta-1)(\omega+1)
t^{2\beta-3}+B\beta(\beta-1)(\beta-2)t^{\beta-3}}{(\omega+1)V_{0}+\frac{\omega-1}{2}}
\end{equation}

In this case using (16), (17), (28), (29), (30) and (31) we get,
\begin{equation}
\epsilon=\frac{2(\beta-1)^2((\omega+1)V_{0}+\frac{\omega-1}{2})}{3B^2
\beta^2(\omega+1) t^{2\beta}+2B\beta(\beta-1)t^{\beta}}
\end{equation}and
\begin{equation}
\eta=\frac{(\beta-1)(\beta-2)(2(\omega+1)V_{0}+\omega-1)}{B\beta
t^\beta(3B\beta(\omega+1) t^{\beta}+2\beta-2)}
-\frac{(\beta-1)^2(2(\omega+1)V_{0}+\omega-1)(3B\beta(\omega+1)t^\beta+\beta-2)}{B\beta
t^\beta(3B\beta(\omega+1)t^\beta+2\beta-2)^2}
\end{equation}

Fig.4 ~represents the variation of $\eta$ against $\epsilon$ for
different values of $\beta$. From the figure, it has been seen that $\eta$
decreases with $\epsilon$ with $\eta$ is always positive and $\epsilon$ is always negative.\\
\begin{figure}
\includegraphics[height=3in]{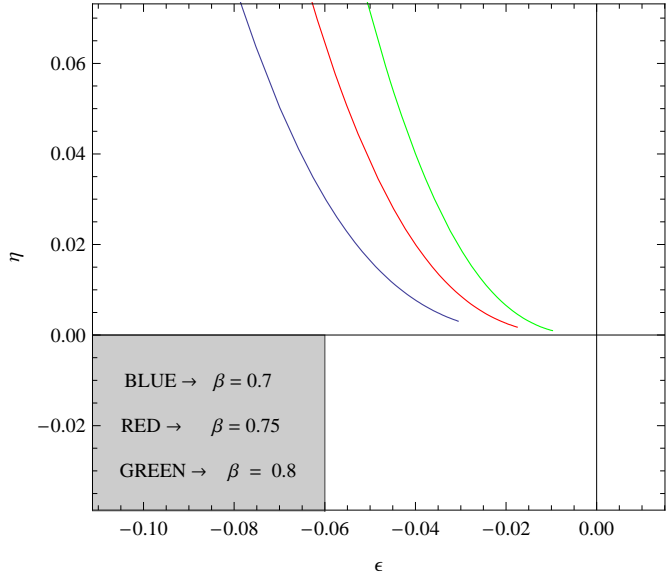}~~~~ \caption{The
variation of $\eta$ against $\epsilon$ from (32) and (33) for $B =
3$,$\omega$=0.2,$V_{0}$=2 and $\beta$=0.7,0.75,0.8} \vspace{7mm}
\end{figure}

Thus from (12) and (14) the energy density becomes.
\begin{equation}
\rho=\rho_{0}a^{-3(1+\omega)}f^\omega=\rho_{0}\exp(-3B(1+\omega)t^\beta)f^\omega
\end{equation}
From (11) and (32) the pressure becomes,
\begin{equation}
p=\rho_{0}\omega a^{-3(1+\omega)}f^\omega=\rho_{0}\omega
\exp(-3B(1+\omega)t^\beta) f^\omega
\end{equation}
Now from (5), (8), (15), (32) and (33) we have,
\begin{equation}
(2V_{0}+1)\dot{\phi}\ddot{\phi}+3B\beta
t^{\beta-1}\dot{\phi}^2=-\rho_{0}(1+\omega)\exp(-3B(1+\omega)t^\beta)f^\omega
\dot{f}
\end{equation}

So, equation (34), (35) and (36) together gives,
\begin{eqnarray*}
\rho_{0}(1+\omega)f^\omega\dot{f}=-\frac{\exp(3B(1+\omega)t^\beta)}{(\omega+1)V_{0}+\frac{\omega-1}{2}}\times
\end{eqnarray*}
\begin{equation}
[B\beta t^{\beta-3}(\beta-1)(2V_{0}+1)(3B \beta(\omega+1)
t^{\beta}+\beta-2)+3B^2\beta^2 t^{2\beta-3}(3B\beta(\omega+1)
t^\beta+2\beta-2)]
\end{equation}
 Integrating both sides of the above equation w.r.t. $t$ and
after a further calculation we obtain,
\begin{equation}
f=\left[-\int\left\{\frac{\exp(3B(1+\omega)t^\beta)}{(\omega+1)V_{0}+\frac{\omega-1}{2}}\times
\frac{x}{\rho_{0}}\right\}~ dt\right]^\frac{1}{1+\omega}
\end{equation}
where, $x=B\beta t^{\beta-3}(\beta-1)(2V_{0}+1)(3B \beta(\omega+1)
t^{\beta}+\beta-2)+3B^2\beta^2 t^{2\beta-3}(3B\beta(\omega+1)
t^\beta+2\beta-2)$.\\

So from equation (32), (33) and (38) we get the expressions of
energy density and pressure as
\begin{equation}
\rho=\rho_{0} \exp(-3B\beta
(1+\omega)t^\beta)\left[-\int\frac{\exp(3B(1+\omega)t^\beta)}{(\omega+1)V_{0}+\frac{\omega-1}{2}}\times
\frac{x}{\rho_{0}}~dt\right]^\frac{\omega}{1+\omega}
\end{equation}
and
\begin{equation}
p=\rho_{0}\omega \exp(-3B\beta
(1+\omega)t^\beta)\left[-\int\frac{\exp(3B(1+\omega)t^\beta)}{(\omega+1)V_{0}+\frac{\omega-1}{2}}
\times\frac{x}{\rho_{0}}~dt\right]^\frac{\omega}{1+\omega}
\end{equation}

\subsection{\bf\large{Chameleon Field in case of Logamediate
Inflation:-}}
 Consider a particular
scenario of {\bf Logamediate Inflation} [14], where the scale
factor $a(t)$ is described as, \begin{equation} a(t)=\exp (A(\ln
t)^{\alpha})
\end{equation}
 where $A \alpha>0$ and $\alpha>1$. The Hubble parameter $H=\frac{\dot{a}}{a}$
  becomes,
\begin{equation}
H=\frac{A\alpha}{t}(\ln t)^{\alpha-1}
\end{equation}
Hence from (42) we get,
\begin{equation}
\frac{\dot H}{H}=\frac{\alpha-1-\ln t}{t\ln t}
\end{equation}
and
\begin{equation}
\frac{\ddot H}{H}=\frac{2(\ln t)^2-3(\alpha-1)\ln
t+(\alpha-1)(\alpha-2)}{t^2 (\ln t)^2}
\end{equation}

From (41) we get,
\begin{equation}
\frac{\ddot{a}}{a}=\frac{A^2
\alpha^2}{t^2}(\ln{t})^{2\alpha-2}-\frac{A\alpha}{t^2}(\ln{t}-\alpha+1)(\ln{t})^{\alpha-2}
\end{equation}

 Putting the value of $a(t)$ the declaration
parameter $q=-\frac{a \ddot{a}}{\dot{a}^2}$ we get,
\begin{equation}
q=-1+\frac{\ln t-\alpha+1}{A\alpha (\ln{t})^\alpha}
\end{equation}

Fig.5 represents the variation of $q$ against $H$ for different
values of $\alpha$.\\

\begin{figure}
\includegraphics[height=2in]{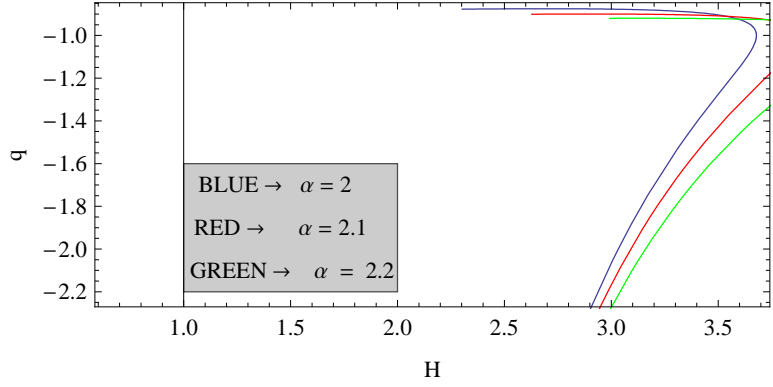}~~~~ \caption{The
variation of $q$ against $H$ from (46) for $A =
1$,$\omega=.2$,$V_{0}$=2 and $\alpha=2,2.1,2.2$} \vspace{7mm}
\end{figure}

From (20), (41), (42) and (46) we get the expressions for the
statefinder parameters as
\begin{equation}
r=1+\frac{3(\alpha-1)}{A\alpha (\ln{t})^\alpha}-\frac{3}{A\alpha
(\ln{t})^{\alpha-1}}+\frac{2}{A^{2}\alpha^{2}(\ln{t})^{2\alpha-2}}
-\frac{3(\alpha-1)}{A^{2}\alpha^{2}(\ln{t})^{2\alpha-1}}
+\frac{(\alpha-1)(\alpha-2)}{A^2\alpha^2 (\ln{t})^{2\alpha}}
\end{equation}
and
\begin{equation}
s=\frac{\frac{3(\alpha-1)}{A\alpha(\ln{t})^{\alpha}}-\frac{3}{A\alpha
(\ln{t})^{\alpha-1}}+\frac{2}{A^{2}\alpha^{2}(\ln{t})^{2\alpha-2}}
-\frac{3(\alpha-1)}{A^{2}\alpha^{2}(\ln{t})^{2\alpha-1}}
+\frac{(\alpha-1)(\alpha-2)}{A^{2}\alpha^{2}(\ln{t})^{2\alpha}}}
{\frac{3}{A\alpha(\ln{t})^{\alpha-1}}-\frac{3(\alpha-1)}{A\alpha(\ln{t})^{\alpha}}-\frac{9}{2}}
\end{equation}

Fig.6 represents the variation of $s$ against $r$ for different
values of $\alpha$.\\

\begin{figure}
\includegraphics[height=2.5in]{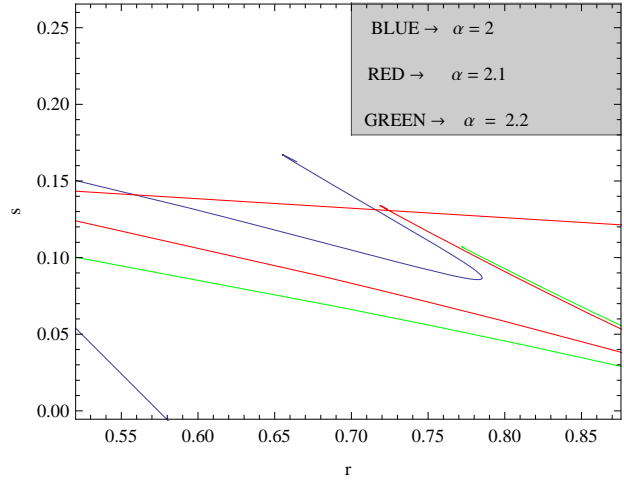}~~~~ \caption{The
variation of $s$ against $r$ from (47) and (48) for $A =
1$,$\omega=.2$,$V_{0}$=2 and $\alpha=2,2.1,2.2$} \vspace{7mm}
\end{figure}

From equation (6), (7), (8), (11), (42) and (45) we get,
\begin{equation}
2 \left\{\frac{A^2
\alpha^2}{t^2}(\ln{t})^{2\alpha-2}-\frac{A\alpha}{t^2}(\ln{t}-\alpha+1)(\ln{t})^{\alpha-2}\right\}
+ (3\omega+1)\left\{\frac{A\alpha}{t}(\ln t)^{\alpha-1}\right\}^2
=\dot{\phi}^2 \left[V_{0}(\omega+1)+\frac{1}{2}(\omega-1)\right]
\end{equation}

Thus using (8) and (49) we get the Chameleon Field and Chameleon
Potential as,

\begin{equation}
\phi=\int\sqrt{\frac{3A^2
\alpha^2(\omega+1)(\ln{t})^{2\alpha-2}-2A\alpha(\ln{t}-\alpha+1)(\ln{t})^{\alpha-2}
}{(\omega+1)V_{0}+\frac{\omega-1}{2}}}~\cdot~\frac{dt}{t}
\end{equation}
and
\begin{equation} V(\phi)=\frac{V_{0}}{t^2}\times\frac{3A^2
\alpha^2(\omega+1)(\ln{t})^{2\alpha-2}-2A\alpha(\ln{t}-\alpha+1)(\ln{t})^{\alpha-2}
}{(\omega+1)V_{0}+\frac{\omega-1}{2}}
\end{equation}

Fig.7 ~represents the variation of $V$ against $\phi$ for
different
values of $A$ and $\alpha$.\\

\begin{figure}
\includegraphics[height=2in]{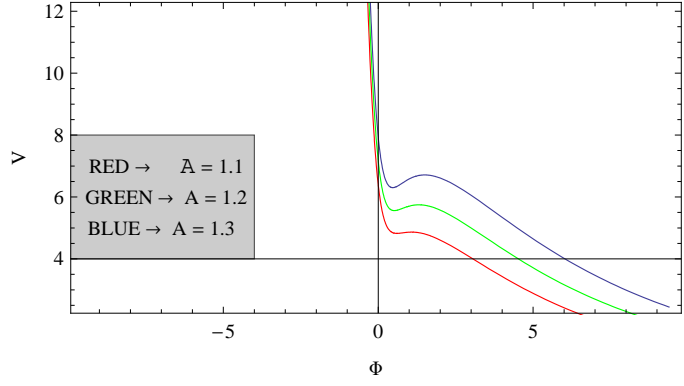}~~~~ \caption{The
variation of $V$ against $\phi$ from (50) and (51) for $A =
1.1,1.2,1.3$,~ $\omega=0.2$,$V_{0}$=2 and $\alpha=2$}\vspace{7mm}
\end{figure}

From (28), (29), (43), (44) and (50) we get the slow-roll
parameters,
\begin{equation}
\epsilon=\frac{2(\ln{t}-\alpha+1)^2\times((\omega+1)V_{0}+\frac{\omega-1}{2})}
{3A^2 \alpha^2(\omega+1)(\ln{t})^{2\alpha}
-2A\alpha(\ln{t}-\alpha+1)(\ln{t})^\alpha}
\end{equation}and
\begin{eqnarray*}
\eta=\frac{(2V_{0}(\omega+1)+\omega-1)\times(2(\ln
t)^2-3(\alpha-1)\ln{t}+(\alpha-1)(\alpha-2))}{3A^2
\alpha^2(\omega+1)(\ln{t})^{2\alpha}
-2A\alpha(\ln{t}-\alpha+1)(\ln{t})^\alpha}+\frac{((\omega+1)V_{0}+\frac{\omega-1}{2})(\ln
t-\alpha+1)}{A\alpha(\ln t)^{\alpha-2}}\times
\end{eqnarray*}
\begin{equation}
\frac{-3A\alpha(\omega+1)(\ln{t}-\alpha+1)(\ln{t})^\alpha+(2(\ln
t)^2-3(\alpha-1)\ln{t}+(\alpha-1)(\alpha-2))}{[3A\alpha(\omega+1)(\ln
t)^\alpha-2(\ln t-\alpha+1)]^2}
\end{equation}

Fig.8 ~represents the variation of $\eta$ against $\epsilon$ for
different values of $\alpha$. It has been seen that $\eta$ is decreases
with $\epsilon$.\\

\begin{figure}
\includegraphics[height=2.5in]{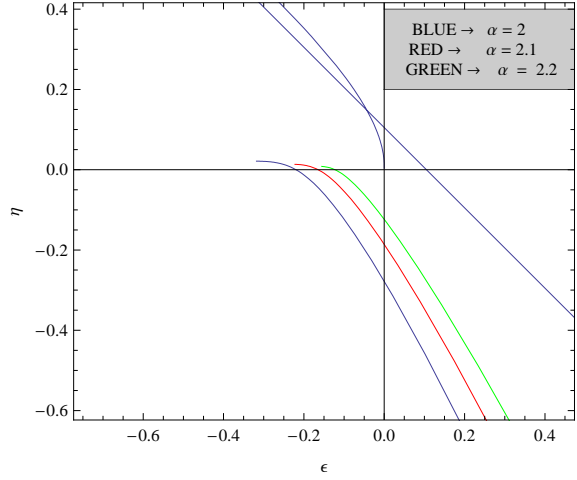}~~~~ \caption{The
variation of $\eta$ against $\epsilon$ from (52) and (53) for $A =
1$,$\omega$=0.2,$V_{0}$=2 and $\alpha$=2,2.1,2.2} \vspace{7mm}
\end{figure}

Thus from (11), (12) and (41) the energy density and pressure
becomes.
\begin{equation}
\rho=\rho_{0}a^{-3(1+\omega)}f^\omega=\rho_{0}\exp(-3A(1+\omega)(\ln{t})^\alpha)f^\omega
\end{equation}
and
\begin{equation}
p=\rho_{0}\omega a^{-3(1+\omega)}f^\omega=\rho_{0}\omega
\exp(-3A(1+\omega)(\ln{t})^\alpha) f^\omega
\end{equation}
Now from (5), (8), (42), (54) and (55) we have,
\begin{equation}
(2V_{0}+1)\dot{\phi}\ddot{\phi}+\frac{3A\alpha}{t}(\ln{t})^{\alpha-1}\dot{\phi}^2=-\rho_{0}(1+\omega)\exp(-3A(1+\omega)(\ln{t})^\alpha))f^\omega
\dot{f}
\end{equation}

From (50) we get,
\begin{equation}
{\dot\phi}^2=\frac{3A^2
\alpha^2(\omega+1)(\ln{t})^{2\alpha-2}-2A\alpha(\ln{t}-\alpha+1)(\ln{t})^{\alpha-2}
}{t^2((\omega+1)V_{0}+\frac{\omega-1}{2})}
\end{equation}

Differentiate the above equation w.r.t. $t$ we get,
\begin{equation}
\dot{\phi}\ddot{\phi}=\frac{A\alpha(\ln
t)^{\alpha-3}}{((\omega+1)V_{0}+\frac{\omega-1}{2})t^3}[-3A\alpha(\omega+1)(\ln{t}-\alpha+1)(\ln{t})^\alpha+(2(\ln
t)^2-3(\alpha-1)\ln{t}+(\alpha-1)(\alpha-2))]
\end{equation}

So, equations (5), (8), (11), (57) and (58) together give,
\begin{eqnarray*}
\rho_{0}(1+\omega)f^\omega\dot{f}\exp(-3A(1+\omega)(\ln{t})^\alpha)=-\frac{3A^2\alpha^2(\ln
t)^{2\alpha-3}}{((\omega+1)V_{0}+\frac{\omega-1}{2})t^3}[3A\alpha(\omega+1)(\ln
t)\alpha-2(\ln t-\alpha+1)]
\end{eqnarray*}
\begin{equation}
-\frac{(2V_{0}+1)A\alpha(\ln
t)^{\alpha-3}}{((\omega+1)V_{0}+\frac{\omega-1}{2})t^3}[-3A\alpha(\omega+1)(\ln{t}-\alpha+1)(\ln{t})^\alpha+(2(\ln
t)^2-3(\alpha-1)\ln{t}+(\alpha-1)(\alpha-2))]
\end{equation}
Assume,   $X(t)=-\frac{3A^2\alpha^2(\ln
t)^{2\alpha-3}}{((\omega+1)V_{0}+\frac{\omega-1}{2})t^3}[3A\alpha(\omega+1)(\ln
t)\alpha-2(\ln t-\alpha+1)]$\\  and
$Y(t)=-\frac{(2V_{0}+1)A\alpha(\ln
t)^{\alpha-3}}{((\omega+1)V_{0}+\frac{\omega-1}{2})t^3}[-3A\alpha(\omega+1)(\ln{t}-\alpha+1)(\ln{t})^\alpha+(2(\ln
t)^2-3(\alpha-1)\ln{t}+(\alpha-1)(\alpha-2))]$\\\\

 Integrating both sides of the above equation w.r.t. $t$ and
after a further calculation we obtain
\begin{equation}
f=\left[\frac{1}{\rho_{0}}\int{\exp(3A(1+\omega)(\ln{t})^\alpha)(X(t)+Y(t))}dt\right]^\frac{1}{1+\omega}
\end{equation}

So from equations (54), (55) and (60) we get the expressions of
the energy density and pressure as
\begin{equation}
\rho=\rho_{0}
\exp(-3A(1+\omega)(\ln{t})^\alpha)\left[\frac{1}{\rho_{0}}\int{\exp(3A(1+\omega)(\ln{t})^\alpha)(X(t)+Y(t))}dt\right]^\frac{\omega}{1+\omega}
\end{equation}
and
\begin{equation}
p=\rho_{0}\omega\exp(-3A(1+\omega)(\ln{t})^\alpha)
\left[\frac{1}{\rho_{0}}\int{\exp(3A(1+\omega)(\ln{t})^\alpha)(X(t)+Y(t))}dt\right]^\frac{\omega}{1+\omega}
\end{equation}

\section{\normalsize\bf{Discussions}}

In this work, we have considered a model of the flat FRW universe
filled with cold dark matter and Chameleon field where the scale
function is taken as, (i) Intermediate Expansion and (ii)
Logamediate Expansion. In the both cases we find the expressions
of Chameleon field, Chameleon potential, statefinder parameters
and slow-roll parameters. In both the cases, it has been shown
that the potential is always decreases with the chameleon field.
From figures, we have seen the nature of slow-roll parameters
i.e., $\eta$ is always decreasing with $\epsilon$.  We have taken
some particular values of the parameters and constants for the
graphical representation. In both the cases, we have found the
expressions of $f$, energy density and pressure of the cold dark
matter in terms of cosmic time $t$.\\

{\bf References:}\\
\\
$[1]$  N. A. Bachall, J. P. Ostriker, S. Perlmutter and P. J.
Steinhardt, {\it Science} {\bf 284} 1481 (1999).\\
$[2]$ S. J. Perlmutter et al, {\it Astrophys. J.} {\bf 517} 565
(1999).\\
$[3]$ A. G. Riess et al, {\it Astron. J.} {\bf 116} 1009
(1998).\\
$[4]$ P. Brax and J. Martin, astro-ph/0210533.\\
$[5]$ J. Khoury and A. Weltman, {\it Phys. Rev. Lett.} {\bf 93} 171104 (2004); arXiv:astro-ph/0309300.\\
$[6]$ S. Perlmutter et al., {\it Bull. Am. Astron. Soc.} {\bf 29}
1351 (1997).\\
$[7]$ J. L. Tonry et al., {\it Astrophys. J.} {\bf 594} 1 (2003).\\
$[8]$ S. Bridle, O. Lahav, J. P. Ostriker and P. J. Steinhardt,
{\it Science} {\bf 299} 1532 (2003).\\
$[9]$ P. Brax, C. van de Bruck, A. C. Davis, J. Khoury and A.
Weltman, {\it AIP Conf. Proc.} {\bf 736} 105 (2005); astro-ph/0410103.\\
$[10]$ H. Wei and R. -G. Cai, {\it Phys. Rev. D} {\bf 71} 043504 (2005).\\
$[11]$ D. F. Mota and D. J. Shaw, {\it Phys. Rev. Lett.} {\bf 97} 151102 (2006).\\
$[12]$ N. Banerjee, S. Das and K. Ganguly, arXiv:0801.1204v1[gr-qc].\\
$[13]$ N. Banerjee and S. Das, {\it Phys. Rev. D} {\bf 78} 043512 (2008).\\
$[14]$ J. D. Barrow and N. J. Nunes, {\it Phys. Rev. D} {\bf 76}
043501 (2007).\\
$[15]$ V. Sahni, T. D. Saini, A. A. Starobinsky and U. Alam, {\it
JETP Lett.} {\bf 77} 201 (2003).\\

\end{document}